\documentclass[11pt]{article}
\usepackage{geometry}
\usepackage[parfill]{parskip}
\usepackage{amssymb}
\usepackage{amsthm}
\usepackage{amsmath}
\usepackage{latexsym}
\usepackage{graphicx}
\usepackage{color}
\usepackage[applemac]{inputenc}
\usepackage[OT1]{fontenc}
\DeclareGraphicsExtensions{.eps}

\newtheorem{defi}{Definition}[section]
\newtheorem{prop}{Proposition}[section]

\newcommand{\bea}{\begin{eqnarray}}
\newcommand{\ena}{\end{eqnarray}}

\newcommand{\bee}{\begin{enumerate}}
\newcommand{\ene}{\end{enumerate}}

\newcommand{\bei}{\begin{itemize}}
\newcommand{\eni}{\end{itemize}}

\newcommand{\bra}{\begin{array}}
\newcommand{\era}{\end{array}}
\newcommand{\bqn}{\begin{eqnarray}}
\newcommand{\eqn}{\end{eqnarray}}
\newcommand\ben{\begin{enumerate}}
\newcommand\een{\end{enumerate}}

\newcommand{\I}{\mathbb I}

\def\lg{\langle }
\def\rg{\rangle }
\def\llg{\left\langle }
\def\rrg{\right\rangle }
\def\lpa{\left(}
\def\rpa{\right)}

\def\adg{a^{\dag}}

\def\deq{\stackrel{\mathrm{def}}{=}}

\def\Z{\mathbb{Z}}

\def\R{\mathbb{R}}
\def\N{\mathbb{N}}
\def\C{\mathbb{C}}

\def\1{\mathbf {Id} }

\newcommand{\id}{\mathbb I}

\begin{document}
\title{Coherent state quantization of angle, time, and more irregular functions and distributions}
\author{B Chakraborty$^1$, J P Gazeau$^2$, and A Youssef$^2$}
\author{Biswajit Chakraborty$^{\:\mathrm{a}}$, Jean Pierre Gazeau$^{\:\mathrm{b}}$ and Ahmed Youssef$^{\:\mathrm{b}}$
\footnote{e-mail:biswajit@bose.res.in, gazeau@apc.univ-paris7.fr,  youssef@apc.univ-paris7.fr}\\
\\
\emph{$^{\mathrm{a}}$ S. N. Bose National Centre for Basic Sciences,}\\
\emph{JD Block, Sector III, Salt Lake, Kolkata-700098, India}\\
\emph{$^{\mathrm{b}}$ Laboratoire APC,
Universit\'e Paris 7-Denis Diderot,}   \\
\emph{10, rue A. Domon  et  L. Duquet
75205 Paris Cedex 13, France}}
\maketitle
\begin{abstract}
The domain of application of quantization methods is traditionally restricted to smooth classical observables. We show that  the  coherent states  or ``anti-Wick'' quantization enables us to construct fairly reasonable quantum versions of irregular observables living on the classical phase space, such as the angle function, the time function of a free particle  and even a large set of distributions comprising the tempered distributions.
\end{abstract}
\section{Introduction}
\label{intro12}
In this work, we reexamine the way in which Gaussian (or standard) coherent states (CS) allow a natural quantization (``Berezin-Klauder CS or anti-Wick quantization'') of the complex plane $\C = \{z = (q + ip)/\sqrt{2}\}$ viewed as the phase space of the particle motion on the line. First, we extend the  definition of what should be considered as an acceptable quantum observable. Then, we  prove that many classical singular functions give rise to such reasonable quantum operators. More precisely, we  apply the CS  quantization scheme  to classical observables which are not smooth functions or, even more, which are, with mild restrictions,  distributions on the plane. In particular, this departure from the canonical quantization principles allows us to put in a CS diagonal form the argument function $\C \ni z=r e^{i \theta}  \mapsto  \textrm{arg}\: z=\theta$ and the time function of a free particle $\C \ni z  \mapsto  i \frac{z+\bar{z}}{z-\bar{z}}=\cot{\theta}= q/p$. We also consider the Dirac distribution on the plane and its derivatives, and this allows us to reach any kind of finite-dimensional  projector  on the  Hilbert space of quantum states. Finally, we extend this quantization scheme to a set of distributions which includes the space of  tempered distributions.
    
The motivation for enlarging the space of quantizable classical observable also stems from the fact that this coherent state quantization can have possible applications in a wide variety of physical problems, like the long standing and controversial question of the determination and the study of the time operator for an interacting particle (see \cite{timeop} and references therein). This aspect will be considered in this paper in the simplest case of the one-dimensional motion of a free particle (the quantization of $q/p$) or of the harmonic oscillator (angle operator). Our approach has also possible implications in noncommutative (NC) quantum mechanics, which is being  currently studied for its possible  application in fractional Quantum Hall Effect (FQHE): If one considers the Landau problem in a 2D plane, the commutators of the projected $x$ and $y$ coordinate operators of a particle onto the lowest Landau level give rise to noncommutativity in terms of the inverse of the applied magnetic field \cite{szabo}. One is therefore led to study the planar NC quantum mechanics per se, where the ``classical'' Hilbert space itself corresponds to the  Hilbert space of quantum states for the particle motion on the line. The quantum Hilbert space for this planar NC system is thus identified with the set of all bounded operators in this classical Hilbert space, with respect to a  certain inner product \cite{biswa}. One can then introduce a disk \cite{biswa} or defects  \cite{pinzul}  in the NC plane in terms of these projectors in the classical Hilbert space. These defects, on turn,  can give rise to certain edge states, relevant for FQHE.

\section{\label{sec121} The Berezin-Klauder or anti-Wick quantization of the 
motion of a particle on the line}
Let us consider  the quantum motion of a particle  on the real line. On the classical level, the  phase space  (with suitable physical units) reads as $ \C = \{  z = \frac{1}{\sqrt2}(q+ip) \} \cong \R^2$. This phase space is equipped with the ordinary Lebesgue measure on the plane which coincides  with the symplectic 2-form : $\frac{1}{\pi}\,  d^2 z$ where $d^2z =d\Re z\,d\Im z$.  Strictly included in the Hilbert space  $L^2( \C, \frac{1}{\pi}\,  d^2 z)$ of all  complex-valued functions on the complex plane which are square-integrable with respect to this  measure, there is 
the Fock-Bargmann Hilbert subspace $\mathcal{FB}$ of all square integrable functions which are of the form $\phi(z,\bar{z}) = e^{-\frac{\vert z \vert^2}{2}} g(\bar{z})$    where $g(z)$ is analytical entire. As an orthonormal basis of this subspace we have chosen the normalized
powers of the conjugate of the complex variable $z$ weighted by the Gaussian function, \emph{i.e.} $\phi_n (z,\bar{z}) \equiv e^{-\frac{\vert z \vert^2}{2}}\,\frac{{\bar{z}}^n}{\sqrt{n!}}$ with $n \in \N$. Normalized coherent states are well known \cite{Sch26,klau0,glauber0,sud} and read as the following superposition of number eigenstates:
\begin{equation}| z\rangle =  \sum_n  \overline{\phi_n (z,\bar{z})} | n\rangle =e^{-\frac{\vert z \vert^2}{2}} \sum_{n\in \N}  \frac{z^n}{\sqrt{n!}}| n\rangle\,, \quad \lg z | z\rg = 1\, .  
\label{scs12}
\end{equation}
We here recall   one fundamental feature of the states (\ref{scs12}), namely the resolution of the unity in the Hilbert space $\mathcal{H}$ having as orthonormal basis the set of $|n\rg$:
\begin{equation}
\frac{1}{\pi}\int_{\C}  | z\rangle \langle z| \, d^2 z= \I_{{\mathcal H}}\, .
\label{pscs12}
\end{equation}
The   property (\ref{pscs12}) is crucial for our purpose in setting the bridge between the classical and the quantum world. It encodes  the  quality of  coherent states of being \emph{canonical quantizers} \cite{Berezin} along a guideline established by Klauder and Berezin (and also Toeplitz on a more abstract mathematical level). This \emph{Berezin-Klauder-Toeplitz (BKT) (or anti-Wick, or anti-normal) coherent states quantization},  called hereafter CS quantization, consists in associating with any classical observable $f$, that is a (usually supposed smooth, but we will not retain here this too restrictive attribute) function of phase space variables $(q, p)$ or equivalently of $(z,\bar{z})$, the operator-valued integral
\begin{equation}
\frac{1}{\pi}\int_{\C} f( z, \bar{z})\, | z\rangle \langle z| \, d^2 z = A_f\, .
\label{quantizer}
\end{equation}
The resulting operator $A_f$, if it exists, at least  in a weak sense, acts on the Hilbert space ${\mathcal H}$.  It is worthy to  be more explicit about  what we mean by ``weak sense'': the  integral
\begin{equation}
\label{wksssymb}
 \int_{\C}f(z,\bar{z})\vert \lg \psi | z\rangle\vert^2  \, \frac{d^2 z}{\pi} =\lg \psi | A_f| \psi \rg\, , 
\end{equation}
should be finite for any $|\psi\rg \in \mathcal{H}$ (or $\in$ some dense subset in $\mathcal{H}$). One notices that  if $\psi$ is normalized then (\ref{wksssymb}) represents the mean value of the function $f$ with respect to the $\psi$-dependent probability distribution $z \mapsto \vert \lg \psi | z\rangle\vert^2$ on the phase space.

More mathematical  rigor is necessary here, and we will adopt the following acceptance criteria for a function (or distribution) to belong to the class of quantizable classical observables.

\begin{defi}
\label{defobscs}
A  function $\C \ni z \mapsto f(z,\bar{z}) \in \C$ and more generally a distribution $T \in  \mathcal{D}'(\R^2)$ is a \emph{CS quantizable classical observable} along the  map $f \mapsto A_f$ defined by  (\ref{quantizer}), and more generally by  $T \mapsto A_T$,
\begin{itemize}
\item if the map $\C \ni z = \frac{1}{\sqrt{2}} (q + ip) \equiv (q,p) \mapsto \lg z  | A_f | z \rg$ (resp. $\C \ni z \mapsto \lg z  | A_T | z \rg$) is a smooth ($\sim \in C^{\infty}$) function with respect to the $(q,p)$ coordinates of the phase plane.
\item and, if we restore the dependence on $\hbar$ through $z \rightarrow \dfrac{z}{\sqrt{\hbar}}$, we must get the right semi-classical limit, which means that $\lg \frac{z}{\sqrt{\hbar}}  | A_f | \frac{z}{\sqrt{\hbar}} \rg \approx f(\frac{z}{\sqrt{\hbar}},\frac{\bar{z}}{\sqrt{\hbar}}) \textrm{  as } \hbar \rightarrow 0$. The same asymptotic behavior must hold in a distributional sense if we are quantizing distributions.  
\end{itemize}
\end{defi}

The function $f$ (resp. the distribution $T$)  is  an \emph{upper} or \emph{contravariant} symbol of the operator $A_f$ (resp. $A_T$), and the mean value  $\lg z  | A_f | z \rg$ (resp. $\lg z  | A_T | z \rg$) is the \emph{lower} or \emph{covariant} symbol of the operator $A_f$ (resp. $A_T$). The map $f \mapsto A_f$ is linear and associates with the function $f(z) = 1$ the identity operator in $\mathcal{H}$.
Note that the lower symbol of the operator $A_f$ is the Gaussian convolution of the function $f(z,\bar{z})$: 
\begin{equation} 
\llg \frac{z}{\sqrt{\hbar}}  | A_f | \frac{z}{\sqrt{\hbar}} \rrg= \int \frac{d^2z'}{\pi \hbar } e^{-\frac{\left|z-z'\right|^2}{\hbar}} f\lpa\frac{\bar z'}{\sqrt{\hbar}},\frac{z'}{\sqrt{\hbar}}\rpa. 
\label{gaussconv}
\end{equation}
This expression is of great importance and is actually the reason behind the robustness of CS quantization, since it is well defined for a very large class of non smooth functions and even for a  class of distributions comprising the tempered ones. Equation (\ref{gaussconv}) illustrates nicely the regularizing role of  quantum mechanics  versus classical singularities. 
Note also that the Gaussian convolution helps to carry out  the semi-classical limit, since the latter can be extracted by using a saddle point approximation. For regular functions for which $A_f$ exists, the application of the saddle point approximation is trivial and we have
\begin{equation}
\llg \frac{z}{\sqrt{\hbar}}  | A_f | \frac{z}{\sqrt{\hbar}} \rrg \approx f\lpa\frac{z}{\sqrt{\hbar}},\frac{\bar{z}}{\sqrt{\hbar}}\rpa \textrm{  as } \hbar \rightarrow 0
\end{equation}
For singular functions the semi-classical limit is less obvious and has to be verified for each special case, something we will do systematically for those ones considered in the following sections.
Also, this particular aspect of CS quantization can be very useful in the context of
the quantum mechanical problem of particles moving in the NC plane, as we had mentioned earlier \cite{biswa}. Since in this context the quantum Hilbert space
comprises  the bounded operators in the classical Hilbert space, one can recover the usual coordinate space wave function by taking expectation
values of these operators in the coherent state family (\ref{scs12}), i.e. by obtaining the corresponding lower symbol \cite{gracia}.

Now let us make the CS quantization program more explicit. Expanding bras and kets in (\ref{quantizer}) in terms of the Fock states yields the expression of the operator $A_f$ in terms of its infinite matrix elements $(A_f)_{nn'}\deq \lg n |A_f | n'\rg$:
\begin{equation}
\label{matelAf}
A_f = \sum_{n,n' \geq 0} (A_f)_{nn'} |n\rg\lg n'|\, , \quad (A_f)_{nn'}=\frac{1}{\sqrt{n!n'!}}\, \int_{\C}\frac{d^2 z }{\pi}\, e^{-\vert z \vert^2}\, z^n\, \bar{z}^{n'}\, f(z,\bar{z})\, .
\end{equation} 
In the case where the classical observable is ``isotropic'', i.e. $f(z) \equiv h(\vert z \vert^2)$, then $A_f$ is diagonal, with matrix elements given by a kind of gamma transform:
\begin{equation}
\label{matelAfdis}
(A_f)_{nn'}=\delta_{nn'}\frac{1}{n!}\, \int_{0}^{\infty}du \,  e^{-u}\, u^n\, h(u)\, 
\end{equation}
In the case where the classical observable is purely angular-dependent, i.e. $f(z) = g(\theta)$ for $z = \vert z \vert \, e^{i\theta}$, the matrix elements $(A_f)_{nn'}$ are obtained through a Fourier transform:
\begin{equation}
\label{matelAfdang}
(A_f)_{nn'}=\frac{\Gamma(\frac{n+n'}{2}+1)}{\sqrt{n!n'!}}\, c_{n'-n}(g)\, ,
\end{equation}
where $c_m(g) \deq \frac{1}{2\pi}\, \int_0^{2 \pi} g(\theta)\, e^{-im\theta}\, d\theta$ is the Fourier coefficient of the $2 \pi$-periodic function $g$.
Thus we have in this case:
\begin{equation}
\label{angdep}
A_f=\sum_{n=0}^\infty c_0 \left|n\right\rangle \left\langle n\right| +  \sum_{q=1}^\infty\sum_{n=0}^\infty \frac{\Gamma(\frac{2n+q}{2}+1)}{\sqrt{n!(n+q)!}}\, \Big[c_q \left|n\right\rangle \left\langle n+q\right| + c_{-q} \left|n+q\right\rangle \left\langle n\right|   \Big]\, .
\end{equation}
Let us explore what this quantization map produces starting with some elementary functions $f$. We  have for the most basic one,
\begin{equation}
\int_{\C}  z\, | z\rangle \langle z| \, \frac{d^2 z }{\pi}= \sum_n \sqrt{n+1} 
| n\rangle \langle n+1| \equiv a\, 
\label{low}
\end{equation}
which is the lowering operator, $a | n\rangle = \sqrt{n} | n - 1\rangle$. 
The  adjoint
$a^{\dagger}$ is obtained by replacing $z$ by $\bar{z}$ in (\ref{low}).
From $q = \frac{1}{\sqrt{2}}(z +
\bar{z})$ et  $p = \frac{1}{\sqrt{2}i}(z - \bar{z})$, one easily infers by linearity that the canonical position $q$ and momentum  $p$
map to the quantum observables $\frac{1}{\sqrt{2}}(a + a^{\dagger}) \equiv Q$ and $\frac{1}{\sqrt{2}i}(a - a^{\dagger})
\equiv P$ respectively. In consequence,  the self-adjoint operators $Q$ and $P$ obtained in this way obey the canonical
commutation rule $\lbrack Q, P \rbrack = i\I_{{\mathcal H}}$, and for this reason fully deserve  the name of position and momentum operators of the usual (galilean) quantum mechanics, together with all  localization properties specific to the latter. 

\section{\label{sec122} Canonical quantization rules}

At this point, it is worthy to recall  what  \emph{quantization of classical mechanics} does mean in a commonly accepted sense (for a recent review see \cite{algen}). In this context, a classical observable $f$ is supposed to be a smooth function with respect to the canonical variables. In the above we have chosen units such that the Planck constant is just put equal to 1. Here we reintroduce it since it parametrizes the link between classical and quantum mechanics.
\subsection*{\bf Van Hove canonical quantization rules \cite{van}}
Given a phase space with canonical coordinates $({\pmb q}, {\pmb p})$
\bei
 \item[(i)] to the classical observable $f({\pmb q}, {\pmb p}) = 1$ corresponds the identity operator  in the (projective) Hilbert space ${\mathcal H}$ of quantum states,
\item[(ii)] the correspondence that assigns to a classical observable $f({\pmb q}, {\pmb p})$,  a self-adjoint operator on ${\mathcal H}$ is a linear map,
\item[(iii)] to the classical Poisson bracket corresponds, \underline{at least at the order} $\hbar$, the quantum commutator, multiplied by $i \hbar$:
\begin{align*}
\label{vanhove}
\mathrm{with} \ f_j({\pmb q}, {\pmb p})   & \mapsto A_{f_j} \ \mathrm{for}\  j= 1,2,3   \\
\mbox{we have} \ \left\{ f_1,f_2  \right\} = f_3  & \mapsto  \lbrack A_{f_1}, A_{f_2} \rbrack = i \hbar A_{f_3} + o(\hbar)
\end{align*}
\item[(iv)] some conditions of minimality on the resulting observable algebra.
\eni  
The last point can give rise to technical and interpretational difficulties \cite{algen}.

It is clear that points (i) and (ii) are fulfilled with the  CS quantization, the second one at least for  observables obeying fairly mild conditions. In order to better understand the ``asymptotic" meaning of Condition (iii), let us quantize higher degree monomials, starting with
$H = \frac{\hbar}{2} (p^2 + q^2)=\hbar \vert z \vert^2$, the classical harmonic oscillator Hamiltonian. For the latter, we get immediately from (\ref{matelAfdis}):
\begin{equation}
\label{modzsq}
A_{H} = \hbar A_{\vert z \vert^2}=\hbar \sum_{n\geq 0}(n+1)|n\rg\lg n| =\hbar  N + \hbar \I_{{\mathcal H}} 
\end{equation}
where $N = \adg a $ is the number operator. We see on this elementary example that the  CS quantization does not fit exactly with the canonical one, which consists in just replacing $q$ by $Q$ and $p$ by $P$ in the  expressions of the observables $f(q,p)$ and next proceeding to a symmetrization in order to comply with self-adjointness. In fact, the quantum Hamiltonian obtained by this usual canonical procedure is equal to $\hat{H}= \hbar N + \hbar/2 \I_{{\mathcal H}} $. In the present case, there is a shift by  $\hbar/2$ between the  spectrum of  $\hat{H}$ and our coherent state quantized Hamiltonian $A_H$. Actually, it seems that no physical experiment can discriminate between those two spectra
that differ from each other by a simple shift (for a deepened discussion on this point, see for instance \cite{kastrup}), unless one couples the system with gravity  which couples to any system carrying energy and momentum. \footnote{This can be considered, on a quite elementary level,  as a facet  of the cosmological constant problem, since the inclusion of a cosmological constant $\Lambda$ corresponds to a shift in the Hamiltonian $H \rightarrow H+ \int d^3x \Lambda$. See \cite{wein} for a review on this question. 

In the same spirit, Wigner showed in \cite{wig} that  the usual canonical commutation relation $\left[Q,P\right]=i \hbar \I$ is not the only one compatible with the requirement that the quantum operators in the Heisenberg picture obey the classical equations of motion. In fact for the harmonic oscillator (with unit mass and frequency) a whole family of commutation relations parametrized by the ground state energy $E_0$ are admissible:
\begin{equation}
\left(\left[Q_W,P_W\right]-i \I\right )^2=-\left(2 E_0-1\right)^2 \I\, 
\end{equation}      
The canonical commutation relations $\left[Q_W,P_W\right]=i \I$ correspond to $E_0=1/2$. The CS quantization gives $E_0=1$ which would correspond to $\left[Q_W,P_W\right]=-2 i \I$ in the Wigner quantization scheme, and so should entail a (non-canonical!) redefinition of position and momentum, something like $P = Q_W/\sqrt{2}\, , \, Q = P_W/\sqrt{2}$. At this stage, let us recall that  the vacuum energy of a free scalar field of mass $m$ is given by 
$$
\left\langle 0\right|H\left|0\right\rangle=\left\langle 0\right|\int d^3k \left[\omega_k a^\dagger_k a_k + \omega_k E_0\right]\left|0\right\rangle=
E_0 \int d^3k  \sqrt{\vec{k}^2+m^2}
$$  
and it is worth noting that the quantization ambiguity showed by Wigner does not allow $E_0=0$, with all the implications to the cosmological constant problem that such a semi-classical computation would have.}

Let us add for future references the quantization of the Hamiltonian of a free particle moving on the line. The Hamiltonian for the free particle of unit mass is $H(q,p)=\frac{p^2}{2}$. With $z=(q+ i p)/\sqrt{2}=r \textrm{e}^{i \theta}$  the Hamiltonian is $H(z,\bar{z})=r^2 \sin^2 \theta $. Using the expression (\ref{matelAf}) we get the quantum Hamiltonian operator
\begin{equation}
	A_H= \frac{1}{2} \sum_{n=0}^\infty \left[ -\frac{\sqrt{(n+1)(n+2)}}{2}  \Big[\left|n\right\rangle \left\langle n+2\right| +  \left|n+2\right\rangle \left\langle n\right|   \Big] +n \left|n\right\rangle \left\langle n\right|\right]\, .
\end{equation}
We also observe that the lower symbol is exactly equal to the classical Hamiltonian for any value of $z$ 
\begin{equation}
\left\langle z\right|A_H\left|z\right\rangle= r^2 \sin^2 \theta=H(z,\bar{z})\, .
\end{equation}

\section{\label{sec123}More upper and lower symbols: the  angle operator}

\begin{figure}[h]
		\includegraphics[width=0.5\textwidth]{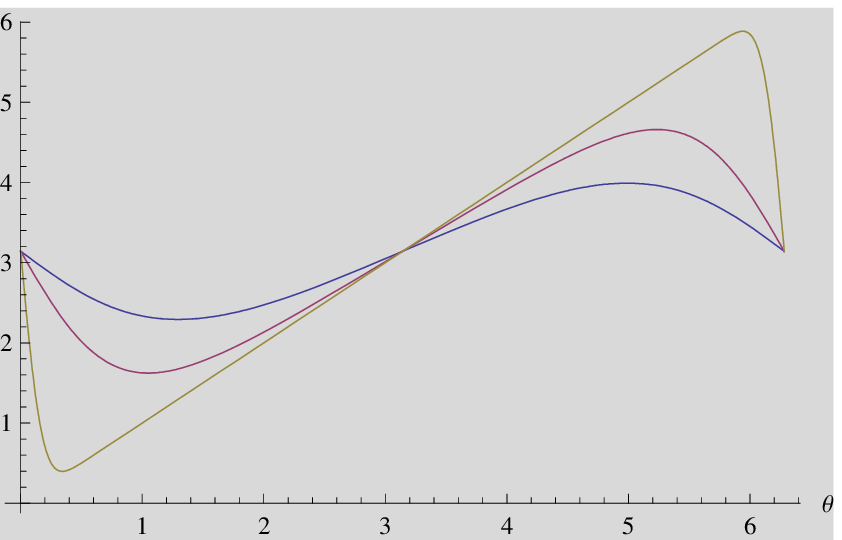}
		\includegraphics[width=0.45\textwidth]{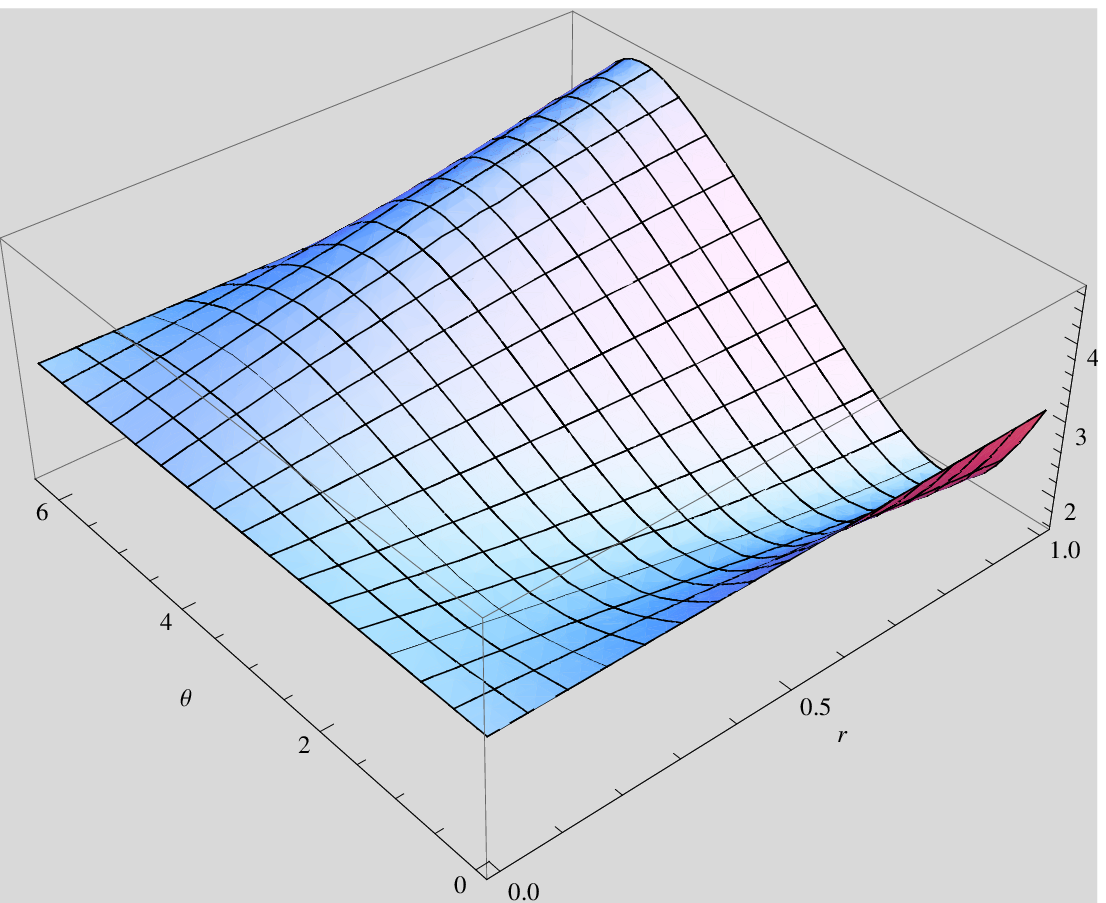}
\caption{Lower symbol of the angle operator for $r = \left\{0.5,1,5\right\}$ and $\theta \in \left[0,2 \pi \right)$  and for $\left(r,\theta \right) \in \left[0,1\right] \times  \left[0,2 \pi \right)$.} 
	\label{lwsymbangle}
\end{figure}

Since we do not retain in our quantization scheme the condition of smoothness on the classical observables, we feel free to CS quantize a larger class of functions on the  plane, like  the argument $\theta \in [0, 2\pi) \, \mathrm{mod}\, 2 \pi$ of the complex variable $z = r\, e^{i\theta}$. The  function $\C \ni z \mapsto \theta = \arg{z}$ is infinite-valued with  a branch cut  starting from the origin which is a branching point. Computing its quantum counterpart from (\ref{matelAfdang}) is straightforward and yields the infinite matrix:
\begin{equation}
\label{scsphaseop}    
A_{\textrm{arg}}= \pi\,  \I_{{\mathcal H}} + i \, \sum_{n\neq n'}\frac{\Gamma\left( \frac{n + n'}{2}+1\right)}{\sqrt{n!n'!}}\, \frac{1}{n'-n}\, |n\rg\lg n'|\, .
\end{equation}
The corresponding lower symbol reads as the  Fourier sine series:
\begin{align}
\label{lwsymphaseop}
\nonumber \lg z|A_{\textrm{arg}}|z\rg \equiv \lg (r,\theta)|A_{\textrm{arg}}|(r,\theta)\rg &= \pi + i\, e^{-\vert z \vert^2}\, \sum_{n\neq n'}\frac{\Gamma\left( \frac{n + n'}{2}+1\right)}{n!n'!}\, \frac{z^{n'}\, \bar{z}^n}{n'-n}\\
&= \pi -  \sum_{q = 1}^{\infty}c_q(r)\, \sin{q\theta}\, ,
\end{align}
where
\begin{eqnarray}
\nonumber c_q (r) &=&  \frac{2 r^q}{q} \, \frac{\Gamma(\frac{q}{2} +1)}{\Gamma(q+1)}\, {}_1F_1(\frac{q}{2} +1;q+1; r^2) e^{-r^2} \\
        &=&	 \frac{\sqrt{\pi} r}{q}    \left(I_{\frac{q-1}{2}}(r^2/2)+I_{\frac{q+1}{2}}(r^2/2)\right) e^{-r^2/2}\, .
\end{eqnarray}
We can also write an integral representation of the lower symbol using the convolution (\ref{gaussconv})
$$
\lg z|A_{\textrm{arg}}|z\rg= \frac{e^{-r^2}}{\pi} \int_0^{2 \pi} \mathrm{d}\phi \: \phi \left[1+\sqrt{\pi} r e^{r^2 \cos^2(\theta-\phi)} \cos{(\theta-\phi) \left\{ 1+\textrm{Erf} \left[r \cos(\theta-\phi) \right]\right\} }\right]\, .
$$
Let us  verify  that this lower symbol is $\mathcal{C}^\infty$ as a function of $r$ and $\theta$ in conformation with our definition \ref{defobscs}. First we  note that 
$$
\frac{\mathrm{d}^n}{\mathrm{d} r^n} c_q(r)= \frac{e^{-r^2/2}}{r^m} \left(P(r,q) I_{n+\frac{q-1}{2}}(r^2/2)+Q(r,q) I_{n+\frac{q+1}{2}}(r^2/2)\right)\, , 
$$
where $P$ and $Q$ are polynomials in the variables $(r,q)$ and $(m,n)$ are positive integers. Then we use the asymptotic formula for large order of the Bessel function \cite{watson}
\begin{equation}
I_\nu(x) \approx \frac{1}{\sqrt{2 \pi \nu}} {\left(\frac{x e}{2 \nu}\right)}^\nu \quad \textrm{for large $\nu$}\, . 
\end{equation}
This makes the series $\sum_{q=1}^\infty \dfrac{\mathrm{d}^n}{\mathrm{d} r^n} \left[c_q(r) \sin(q \theta)\right]$ and $\sum_{q=1}^\infty \dfrac{\mathrm{d}^n}{\mathrm{d} \theta^n} \left[c_q(r) \sin(q \theta)\right]$ absolutely convergent, and thus $\lg z|A_{\textrm{arg}}|z\rg$ is $\mathcal{C}^\infty$ for $r>0 \textrm{ and } \theta \in \R$.
The behavior of the lower symbol (\ref{lwsymphaseop}) is shown in Figure \ref{lwsymbangle}.
It is interesting to evaluate the asymptotic behaviors   of  the function (\ref{lwsymphaseop})  at  small and large $r$ respectively. At small $r$, it oscillates around its average value $\pi$ with amplitude equal to $\sqrt{\pi}r$:
$$
\lg (r,\theta)|A_{\textrm{arg}}|(r,\theta)\rg \approx \pi - \sqrt{\pi} r\,\sin{\theta}\,
$$
 At large $r$, we recover the Fourier series of the $2 \pi$-periodic angle function:
 $$
\lg (r,\theta)|A_{\textrm{arg}}|(r,\theta)\rg \approx \pi - 2\,\sum_{q = 1}^{\infty}\frac{1}{q}\, \sin{q\theta} =  \theta \quad \mbox{for} \quad \theta \in [0, 2\pi)\, 
$$
The latter result can be equally understood in terms of classical limit of 
these quantum objects. Indeed, by re-injecting into our formula physical dimensions, we know that the quantity $\vert z \vert^2 = r^2$ acquires the dimension of an action and should appear in the formulas as divided by the Planck constant $\hbar$. Hence, the limit $r \to \infty $ in  our previous expressions can  also be considered as the classical limit  $\hbar \to 0$.
Since we  have at our disposal the number operator $N= \adg \, a$, which is up to a constant shift the quantization of the classical action, and an angle operator, we can examine their commutator and its lower symbol in order to see to what extent we get something close to the expected canonical value, namely $i \,  \I_{{\mathcal H}}$.
The commutator reads as
\begin{equation}
\label{ comactang}
[A_{\textrm{arg}},N] =  i \, \sum_{n\neq n'}\frac{\Gamma\left( \frac{n + n'}{2}+1\right)}{\sqrt{n!n'!}}\,  |n\rg\lg n'|\,  .
\end{equation}
Its lower symbol is then given by
\begin{equation}
\lg (r,\theta)|[A_{\textrm{arg}},N] |(r,\theta)\rg = i\,  \sum_{q = 1}^{\infty} q c_q(r)\, \cos{q\theta} \equiv i\, \mathcal{C}(r,\theta)\, ,
\end{equation}
with the same $c_q(r)$ as in (\ref{lwsymphaseop}).

At small $r$, the function $\mathcal{C}(r,\theta)$ oscillates around $0$ with amplitude equal to $\sqrt{\pi}r$:
$$
\mathcal{C}(r,\theta) \approx  \sqrt{\pi} r\, \,\cos{\theta}\,
$$
At large $r$, the function $\mathcal{C}(r,\theta)$ tends to the Fourier series $2\,\sum_{q = 1}^{\infty} \cos{q\theta}$ whose convergence has to be understood in the sense of distributions. Applying the Poisson summation formula, we get at $r \to \infty$ (or $\hbar \to 0$) the expected ``canonical'' behavior for $\theta \in [0, 2 \pi)$. The fact that this commutator is not exactly canonical  was  expected since we know from Dirac  \cite{dirac27} about the impossibility to get canonical commutation rules for the quantum versions of the classical canonical pair action-angle.
On a more general level, we know that there exist  such classical pairs for which mathematics -e.g. the Pauli theorem-\cite{pauli58}  prevent the corresponding quantum commutator of being exactly canonical. We will discuss this point in more details when quantizing the time function in the next section. However, in the present case, we obtain in the quasi-classical regime the following asymptotic behavior:
 \begin{equation}
\label{poissonangle}
\lg (r,\theta)|[A_{\textrm{arg}}, N]|(r,\theta)\rg \approx -i + 2 \pi i\sum_{n \in \Z} \delta(\theta - 2 \pi n)\, .
\end{equation}
One can observe that the commutator symbol  becomes  ``canonical'' for $\theta \neq 2 \pi n, \, n \in \Z$. Dirac singularities are located at the discontinuity points of the $2 \pi$ periodic extension of the linear function   $f(\theta) = \theta$ for $\theta \in \left[0,2\pi\right)$. 

\section{Classical and quantum time of the free particle}

The quantization of the time function is, like for the angle,  an old, important, and controversial question \cite{timeop}. Aside from conceptual problems, the basic difficulty encountered in the construction of a quantum time operator is summarized in the called Pauli theorem \cite{pauli58,galapon02,giann02,toller96}: one would expect naively  the time operator $T$ to be conjugated to the Hamiltonian $H$. However if one assumes that $H$ is a bounded from below operator, such a commutation relation $\left[T,H\right]=-i \id$ cannot hold. The Hamiltonian for the free particle, $H(q,p)=p^2/2$, implies $q(t)= p\, t\ $ (up to the addition of a constant). We can invert this relation to get an expression of the classical time as a function on the phase space $t=  q/p$. If we view the phase space as the complex plane by setting $z=(q+ i p)/\sqrt{2}=r \textrm{e}^{i \theta}$, then the classical time function is $t(z,\bar{z})=\cot{\theta}$. Since the time function is only $\theta$ dependent $t(z,\bar{z})=g(\theta)$, its CS quantized version is given by 
(\ref{angdep})
. The coefficients $c_m$ are given by
\begin{align}
\nonumber c_m(g) &= \\
&\frac{1}{2\pi}\, \int_0^{2 \pi} g(\theta)\, e^{-im\theta}\, d\theta=\frac{1}{2\pi}\, \int_0^{2\pi} \cot(\theta) \cos{(m \theta)} + i \frac{1}{2\pi}\, \int_0^{2\pi} \cot(\theta) \sin{(m \theta)}\, .
\end{align}
The first integral is singular and we will understand it as a principal value, for instance, $\int_0^{2 \pi} \cot(\theta)\, d\theta=\lim_{\epsilon\rightarrow 0} \int_\epsilon^{ \pi-\epsilon} \cot(\theta)\, d\theta+\int_{\pi+\epsilon}^{2\pi-\epsilon} \cot(\theta)\, d\theta$. For parity reasons the real part of $c_m$ is zero and
$$
c_m = i \frac{1}{2\pi}\, \int_0^{2\pi} \cot(\theta) \sin{(m \theta)}= \left\{ \begin{array}{ll}
i  & \textrm{if $m>0$ and even}\\
-i  & \textrm{if $m<0$ and even}\\
0 & \textrm{otherwise}\, .
\end{array} \right.
$$
The time operator is thus given by:
\begin{equation}
A_t=i \sum_{k=1}^\infty\sum_{n=0}^\infty \frac{(n+k)!}{\sqrt{n!(n+2k)!}}\, \Big[ \left|n\right\rangle \left\langle n+2k\right| -  \left|n+2k\right\rangle \left\langle n\right|   \Big]\, , 
\end{equation}
and the lower symbol is:
\begin{equation}
\left\langle z\right|A_t\left|z\right\rangle= \sum_{q=1}^\infty c_q(r) \sin(2 q \theta)\, , 
\end{equation}
where\begin{eqnarray}
\nonumber c_q(r) &=& 2 \sqrt{\pi} e^{-r^2} \left({\frac{r}{2}}\right)^{2 q} \frac{_1F_1(1+q,1+2q,r^2)}{\Gamma(q+1/2)} \\
                 &=& 2 \sqrt{\pi} e^{-r^2/2} \left(I_{q-1/2}(r^2/2)+I_{q+3/2}(r^2/2) \right)\, .
\end{eqnarray}
One can also prove that this lower symbol is $\mathcal{C}^\infty$ exactly in the same way we proved that $\left\langle z\right|A_{\mathrm{arg}}\left|z\right\rangle$ is $\mathcal{C}^\infty$. It is also important to control the semi-classical limit, which appears as $r \rightarrow \infty$
\begin{equation}
c_q(r) \approx  2  \qquad \textrm{as $r \rightarrow \infty$} \quad \textrm{ and } \qquad 2 \sum_{q=1}^\infty \sin(2 q \theta)=\cot(\theta) \quad \textrm{for } \theta \neq 2 \pi n, n \in \Z 
\end{equation}

\begin{figure}[h]
\centering
\includegraphics[width=0.6\textwidth]{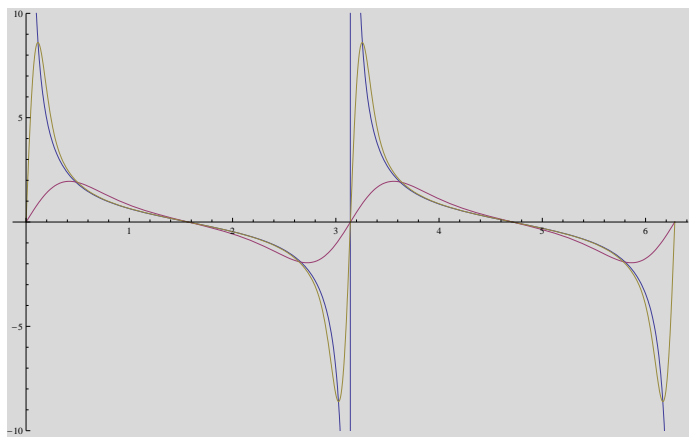}
\caption{The classical time function and the lower symbol of the time operator for $r=2,8$.} 
	\label{timeclassicallimit}
\end{figure}

\subsection{The commutator}
\begin{figure}[h]
\centering
		\includegraphics[width=0.4\textwidth]{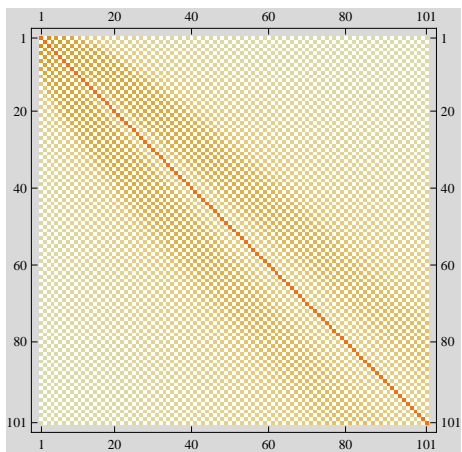}
\caption{Visual representation  of the absolute value of the matrix elements of the commutator truncated to order $100$.} 
	\label{matrixplottimecommutateur}
\end{figure}

\begin{figure}[h]
	\centering
\includegraphics[width=0.5\textwidth]{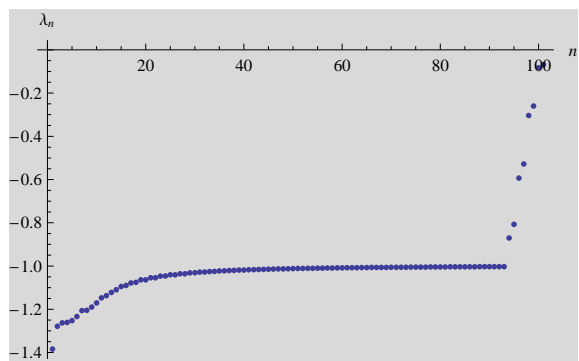}
\caption{Imaginary part of the spectrum of the commutator matrix truncated to order $100$.} 
	\label{spectrecommutateurtemps}
\end{figure}

Using the expressions of the time and Hamiltonian operators, we can compute the commutator $C=\left[A_t,A_H\right]$:
\begin{equation}
\left\langle m\right| \left[A_t,A_H\right]\left|n\right\rangle = \left\{ \begin{array}{ll}
0 & \textrm{if $m-n$ odd}\\
-i & \textrm{if $m=n$ }\\
i \frac{(m-n) \left(\frac{m+n}{2}-1\right)}{4 \sqrt{m! n!}} & \textrm{if $m>n$}\\
-i \frac{(m-n) \left(\frac{m+n}{2}-1\right)}{4 \sqrt{m! n!}} & \textrm{if $m<n$}\, . \\
\end{array} \right.
\end{equation}
The question now is to evaluate the extent to which  this commutator is different from the canonical  $-i \I_{{\mathcal H}}$. First we notice that this matrix shares the same diagonal part as the canonical commutator and is well localized along its diagonal since asymptotically the coefficients are rapidly decreasing  away from the diagonal. For instance, for large $n$, we have $ \left\langle 0\right| \left[A_t,A_H\right]\left|n\right\rangle \approx \textrm{const} \sqrt{(1+12 n) 2^{-n} n^{-{1/2}}}$, which goes to $0$ more rapidly than $e^{-n/3}$. Figure \ref{matrixplottimecommutateur} shows this localization.  
In order to go further in the comparison of our commutator with the canonical one, we numerically study its spectrum by truncating the infinite matrix. The results are shown in figure \ref{spectrecommutateurtemps} and confirms that the spectrum of the commutator $C$ is very close to that of the canonical spectrum $-i$ with infinite degeneracy.

\begin{figure}[h]
\centering
		\includegraphics[width=0.5\textwidth]{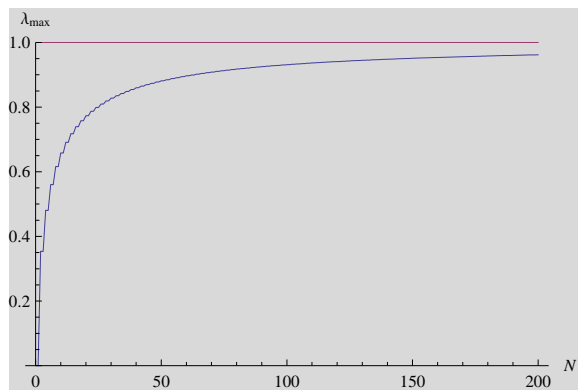}
\caption{Spectral norm  $\lambda_{max}$ of $D=\left[A_t,A_H\right]-(-i) \I_{{\mathcal H}}$ as a function of the truncation order $N$} 
	\label{spectralnormtime}
\end{figure}

To study further the departure from the canonical value of the commutator, let us define the operator $D=\left[A_t,A_H\right]-(-i) \I_{{\mathcal H}}$. First we note that $D$ is a trace class operator, since $\mathrm{Tr} D=0$. We note also that $D$ is not a Hilbert-Schmidt operator because 
$$\mathrm{Tr}\left(D^\dagger D\right)=\frac{1}{2}\sum_{q=1}^\infty \sum_{n=0}^\infty \frac{q^2 \left((n+q-1)!\right)^2}{n! (n+2q)!}$$ is a divergent sum.
We have carried out a  numerical analysis to find the spectrum of the operator $D^\dagger D$ and we find that this spectrum is bounded and seems to verify $\sigma(D^\dagger D) \subset \left[0,1\right]$. Thus the spectral norm of $D$, given by $\sqrt{\mathrm{sup} \  \sigma(D^\dagger D) } $ is well defined. Numerically its value  is equal to 1 as shown in  Figure \ref{spectralnormtime}.    

The lower symbol of the commutator can be written as the following sum
\begin{equation}
\left\langle z\right|\left[A_t,A_H\right]\left|z\right\rangle= -i + i \sum_{q=1}^\infty c_q(r) \cos(2 q \theta)
\end{equation}
where
\begin{equation}
\quad c_q(r)=e^{-r^2} \frac{r^{2q} q!}{(2q)!} \: {}_1F_1\left(q,1+2q,r^2\right)\, .
\end{equation}
Restoring the $\hbar$ units, we can verify that this commutator has the canonical form in the semi-classical limit $\hbar \rightarrow 0$, since in this limit we have $c_q(r) \approx (-1)^q e^{-r^2/\hbar^2} + \hbar^2 \dfrac{q}{r^2}$.

\section{\label{sec124}Quantization of distributions: Dirac and others}
It is commonly accepted that  a ``CS diagonal'' representation of the type (\ref{quantizer})  is possible only for a restricted class of operators in $\mathcal{H}$. The reason is that we usually put too much restrictive conditions on the upper symbol $f(z,\bar{z})$ viewed as a classical observable on the phase space, and so it is submitted to belong to the space of infinitely differentiable functions on $\R^2$. We already noticed that a ``reasonable'' phase or angle operator is easily built starting from the classical discontinuous periodic angle function. We are now going to show that \underline{any} simple projector $\Pi_{nn'} \deq |n\rg\lg n'|$ has also a CS diagonal representation by extending the class of classical observables to distributions on $\R^2$ (for canonical coordinates $(q,p)$ or possibly on $\R^+\times [0, 2 \pi)$ (for $(u \deq r^2,\theta)$ coordinates). Due to the general expression (\ref{matelAf}) for matrix elements of the quantized version of an observable $f$, one can immediately think to  tempered distributions  on the plane only since the functions
\begin{equation}
\label{gauspow}
(z,\overline{z}) \mapsto e^{-\vert z \vert^2}\, z^n\, \bar{z}^{n'}
\end{equation}
are rapidly decreasing $C^{\infty}$ functions on the plane with respect to the canonical coordinates $(q,p)$, i.e. they belong to the Schwartz space $\mathcal{S}(\R^2)$,  or equivalently with respect to the coordinates $(z, \bar{z})$. Actually, we can extend the set of ``acceptable'' observables  to those distributions in $\mathcal{D}'(\R^2)$ which obey the following condition (similar extensions to distributions  have been considered in \cite{sud}, and \cite{bog,bogg} for the Weyl quantization). 

\begin{prop}
\label{defobsdist}
A distribution   $T \in \mathcal{D}'(\R^2)$ is  a \emph{CS quantizable classical observable}
if there exists $\eta < 1$ such that the product $e^{- \eta \vert z \vert^2}\, T \in \mathcal{S}'(\R^2)$, i.e. is a tempered distribution.
\end{prop}

Using complex coordinates is  clearly more convenient and we will adopt the following definitions and notations for tempered distributions. Firstly any function $f(z,\bar{z})$ which is ``slowly increasing" and locally integrable with respect to the Lebesgue measure $d^2 z$ on the plane defines a regular tempered distribution $T_f$, \emph{i.e.} a continuous linear  form on the  vector space $\mathcal{S}(\R^2)$ equipped with the usual topology of uniform convergence at each order of partial derivatives multiplied by polynomial of arbitrary degree \cite{schwartz}.
This definition rests on the map,
\begin{equation}
\label{scdist1}
\mathcal{S}(\R^2) \ni \psi \mapsto \langle T_f,\psi \rangle \deq  \int_{\C}d^2z\, f(z,\bar{z}) \, \psi(z,\bar{z})\, ,
\end{equation}
and the notation is kept for all tempered distributions $T$. 
According to Proposition \ref{defobsdist},  this definition can be extended to locally integrable functions $f(z,\bar{z})$ which increase like $e^{\eta \vert z \vert^2}\, p(z,\bar{z})$ for some $\eta <1$ and some polynomial $p$, and it is easily understood in which way this extends to distributions. Actually, the latter can be characterized as derivatives (in the distributional sense) of such functions.  We recall here that  partial derivatives of distributions  are given by
\begin{equation}
\label{scdist2}
 \llg \frac{\partial^r}{\partial z^r}\, \frac{\partial^s}{\partial \bar{z}^s}\,T,\psi \rrg = (-1)^{r+s}\, \llg T, \frac{\partial^r}{\partial z^r}\, \frac{\partial^s}{\partial \bar{z}^s}\,\psi \rrg \, .
\end{equation}
We also recall that the multiplication of distributions $T$ by smooth functions $\alpha(z,\bar{z}) \in C^{\infty}(\R^2) $ is understood through: 
\begin{equation}
\label{prodscdist}
C^{\infty}(\R^2) \ni \psi \mapsto \langle \alpha T,\psi \rangle \deq \langle  T,\alpha\, \psi \rangle\, .
\end{equation}
Of course, all compactly supported distributions like Dirac and its derivatives,  are tempered and so are \emph{CS quantizable classical observable}. The Dirac distribution supported by the origin of the complex plane is  denoted as usual by $\delta$ (and abusively in the present context by $\delta(z,\bar{z})$) :
\begin{equation}
\label{dirac1}
C^{\infty}(\R^2) \ni \psi \mapsto \langle \delta,\psi \rangle \equiv \int_{\C}d^2z\, \delta (z,\bar{z}) \, \psi(z,\bar{z}) \deq \psi(0,0)\,. 
\end{equation}

Let us now CS quantize the Dirac distribution along the recipe provided by Eqs. (\ref{quantizer}) and (\ref{matelAf}):

\begin{align}
\nonumber \frac{1}{\pi}\int_{\C} \delta( z, \bar{z})\, | z\rangle \langle z| \, d^2 z = & \sum_{n,n' \geq 0} \frac{1}{\sqrt{n!n'!}}\, \int_{\C}\frac{d^2 z }{\pi}\, e^{-\vert z \vert^2}\, z^n\, \bar{z}^{n'}\, \delta(z,\bar{z})\,  |n\rg\lg n'| \\=&\frac{1}{\pi}\,   |z=0\rg\lg z=0|\,=\frac{1}{\pi} \,  \Pi_{00}\, .
\label{quantdir}
\end{align}
We thus find that the ground state (as a projector) is the quantized version of the Dirac distribution supported by the origin of the phase space. The obtention of all possible diagonal projectors $\Pi_{nn} = |n\rg\lg n|$ or even all possible oblique projectors $\Pi_{nn'} = |n\rg\lg n'|$ is based on the quantization of  partial derivatives of the $\delta$ distribution. First let us compute the various derivatives of the Dirac distribution: 
\begin{align}
\nonumber U_{a,b}&=\int_{\C} \,  \left\lbrack\frac{\partial^b}{\partial z^b}\, \frac{\partial^a}{\partial \bar{z}^a}\delta( z, \bar{z})\right\rbrack\, | z\rangle \langle z| \, d^2 z  \\
& = \sum_{n,n' \geq 0} (-1)^{ n + a} \frac{b! \, a!}{(b-n)!}\frac{1}{\sqrt{n!n'!}}\,\delta_{n-b,n'-a} \, \Pi_{nn'}  \, .
\label{quantdirderpow}
\end{align}
Once this quantity $U_{a,b}$ at hand, one can  invert the formula in order to get the oblique projector
$\Pi_{r+s,r}=\left|r+s\right\rangle\left\langle r\right|$  as:
\begin{equation}
\label{oblproj}
\Pi_{r+s,r}=\sqrt{r! (r+s)!} (-1)^s \sum_{p=0}^r \frac{1}{p! (s+p)! (r-p)!} U_{p,p+s}\, ,
\end{equation}
and its upper symbol are given by the distribution supported by the origin:
\begin{equation}
\label{upsymbobl}
f_{r+s,r}(z,\bar{z})=\sqrt{r! (r+s)!} (-1)^s \sum_{p=0}^r \frac{1}{p! (s+p)! (r-p)!} \, \left\lbrack\frac{\partial^{p+s}}{\partial z^{p+s}}\, \frac{\partial^p}{\partial \bar{z}^p}\delta( z, \bar{z})\right\rbrack\, .
\end{equation}
Note that this distribution, as is well known, can be approached, in the sense of the  topology on $\mathcal{D}'(\R^2)$, by smooth functions, like linear combinations of derivatives of Gaussians.  
The diagonal projectors $ \Pi_{r,r}$ are then obtained trivially by setting $s=0$ in (\ref{oblproj}) to get
\begin{equation}
\label{diagproj}
\Pi_{r,r}=\sum_{p=0}^r \frac{1}{p!} \left(
\begin{array}{c}
 r \\
 p
\end{array}
\right) U_{p,p}\, .
\end{equation}
Again in the context of quantum mechanics in the NC plane, one notes that one can define a projection operators $\mathrm{P}=\sum_{r=0}^{N}\Pi_{r,r}$ to
define an analogue of a disk \cite{biswa}. On the other hand, the removal of the ``disk" from the classical Hilbert space defines an analogue of
a defect in the NC plane \cite{pinzul}. 

Using the expressions of the projectors and the linearity of the quantization map $A$, one can formally construct  an inversion (dequantization) operator $A^{-1}$ given by:
\begin{equation}
\label{inversion}
A^{-1}(O)=\sum_{r=0}^\infty \sum_{s=1}^\infty \big[\left\langle r+s\right| O \left| r\right\rangle f_{r+s,r}(z,\bar{z})+ r \leftrightarrow r+s \big] + \sum_{r=0}^\infty \left\langle r\right| O \left| r\right\rangle f_{r,r}(z,\bar{z})
\end{equation}
This inversion map also enables us to construct  a star product $*$ on the classical phase space verifying $A_{f*g}=A_f \: A_g$ (See for instance \cite{hir} for a general review on deformation quantization,and \cite{vo,da,alex,bal} for more  material based on coherent states)
$$
f*g=A^{-1} \left(A_f \: A_g\right)\, . 
$$
Note that this star product involves the upper symbols, in contrast to the Voros star product  \cite{vo,da,alex,bal}, which involves the lower symbols.

Many of the  ideas  around this combination of coherent states with distributions pertain to the domain of Quantum Optics. They are already present in the original works by Sudarshan \cite{sud}, Glauber \cite{glauber1},  Klauder \cite{kl}, Cahill \cite{ca}, Miller \cite{mi} and others. In Quantum Optics the basic idea is that replacing the non diagonal representation of quantum operators (usually, in this context,  one focuses on  the  density operators $\rho$) given by 
$$
A=\int_{\C^2} d^2z_1d^2z_2 \left\langle z_1\right|A\left|z_2\right\rangle \left|z_1\right\rangle\left\langle z_2\right|\,  
$$
by a diagonal one, also called the $P$-representation,
$A=\int_{\C} d^2z P(z, \bar z) \left|z\right\rangle\left\langle z\right|$, can  simplify considerably some calculations. Although this can be considered as the CS quantization of $P(z, \bar z)$, the spirit is quite different since their approach is the inverse of ours: given $A$, then the  question is to find $P(z,\bar z)$. The main results  obtained  in this direction is that one can formally write a $P$-representation for each quantum operator $A$, which is  given by \cite{sud}
\begin{equation}
\label{sudformula}
P(z=r e^{i \theta},\bar z)=\sum_{m,n=0}^\infty \frac{\left\langle n\right|A\left|m\right\rangle \sqrt{n! m!}}{2 \pi r (n+m)!}  e^{r^2+i (m-n) (\theta-\pi)}\, 
\delta^{(m+n)}(r)
\end{equation}
or by  \cite{kl}

\begin{equation}
\label{klformula}
P\left(z=(q+ip)/\sqrt{2}, \bar z\right)=\mathcal{F}^{-1} \left[ \tilde{A}(x,y) \;e^{\frac{x^2+y^2}{2}}\right] \quad \textrm{where} \quad  \tilde{A}(x,y)=\mathcal{F} \left[ \left\langle z\right|A\left|z\right\rangle \right].
\end{equation}
Here $\mathcal{F}$  is the Fourier transform from the $(p,q)$-space  to the $(x,y)$-space, and $\mathcal{F}^{-1}$ is its inverse.
However the question of the validity of such formulas is mathematically non trivial: the convergence in the sense of distributions of (\ref{inversion},\ref{sudformula}) is a difficult problem, and  for instance has  been partially studied  by Miller in \cite{mi}
Manifestly,  the work done in this direction was concentrated on the dequantization problem (finding an associated classical  function to each quantum operator) and this was done in a quite pragmatic spirit in order to simplify computations. Let us note that the existence of such a well-defined dequantization procedure is by no means a physical requirement since the quantum realm is by definition richer than the classical one. A more physical requirement is that the semi-classical limit is well behaved, a property that we have placed at the center of our work.

\section{Application of coherent state formulation in a planar NC system}
As was introduced in \cite{biswa} the classical Hilbert space $\mathcal{H}_{C}$ for a planar noncommutative system satisfying $[\hat x_{i},\hat x_{j}]=i\theta\epsilon_{ij}$ is identified as the boson Fock space 
\begin{equation}
\mathcal{H}_{C}=\mathrm{Span}_{\C}\{|n\rangle\}_{n=0}^{n=\infty};|n\rangle\equiv \frac{(b^{\dagger})^{n}}{\sqrt{n!}}|0\rangle
\label{bfock}
\end{equation}
constructed out of the bosonic creation and annihilation operators $b\equiv\frac{1}{\sqrt{2\theta}}(\hat x_{1}+i\hat x_{2})$ and $b^{\dagger}$ respectively satisfying $[b,b^{\dagger}]=1$. On the other hand the quantum Hilbert space $\mathcal{H}_{Q}$ is identified as the set of bounded operators on $\mathcal{H}_{C}$ :
\begin{equation}
\mathcal{H}_{Q}=\{\psi(\hat x,\hat y):\mathrm{tr}_{\C}(\psi(\hat x,\hat y)^{\dagger}\psi(\hat x,\hat y))<\infty\}
\end{equation}
Here the inner product between any pair of states$|\psi)$ and $|\phi)\in \mathcal{H}_{Q}$ is defined as a trace in the classical Hilbert space $\mathcal{H}_{\C}$
\begin{equation}
(\psi|\phi)=\mathrm{tr}_{\C}(\psi^{\dagger}\phi)
\end{equation}
Note that we are denoting the vectors belonging to $\mathcal{H}_{C}$ and $\mathcal{H}_{Q}$ by $|.\rangle$ and $|.)$ respectively. Thus the coherent state $|z\rangle$ introduced in (\ref{scs12}), which provides an overcomplete system for the quantum Hilbert space of states for a particle moving in a line, now corresponds to the over-complete basis for the classical Hilbert space $\mathcal{H}_{C}$(\ref{bfock}) as well, as these two Hilbert spaces are really isomorphic to each other. Consequently, the inner product can be calculated by using either the countable basis $|n\rangle$ or coherent state family 
\begin{equation}
(\psi,\phi)=(\psi|\phi)=tr_{C}(\psi^{\dagger}\phi)=\sum_{n}\langle n|\psi^{\dagger}\phi|n\rangle =\int{\frac{d^2 z}{\pi}\langle z|\psi^{\dagger}\phi|z \rangle}.
\end{equation}
Using this one can identify the normalized momentum eigenstate as
\begin{equation}
|\vec p)=\sqrt{\frac{\theta}{2\pi}}e^{ip_{i}\hat x_{i}};\,\,  (\vec p^{\;\prime}|\vec p)=\delta^{2}(\vec p^{\;\prime}-\vec p)
\label{delta}
\end{equation}
which are nothing but the operator-valued plane-wave states-a direct generalization from the commutative case. The fact that these plane waves are really the momentum eigenstates can be checked easily by considering the adjoint action of momentum on the state $|\psi)\in \mathcal{H}_{Q}$ [2] 
\begin{equation}
\hat p_{i}^{~q}\psi(\hat x,\hat y)=\frac{1}{\theta}\epsilon_{ij}[\hat x_{j},\psi(\hat x, \hat y)]
\label{xk}
\end{equation}
to get
\begin{equation}
\hat p_{i}^{~q}e^{ip_{k} \hat{x}_{k}}=\frac{1}{\theta}\epsilon_{ij}[\hat x_{j},e^{ip_{k} \hat{x}_{k}}]=p_{i}e^{ip_{k}\hat{x}_{k}}
\label{planewave}
\end{equation}
The adjoint action of the momentum (\ref{xk}) along with left action of the coordinate operator $\hat x_{i}$ on the elements of the quantum Hilbert space $\mathcal{H}_{Q}$
\begin{equation}
\hat x_{i}^{q}\psi(\hat x,\hat y)=\hat x_{i}\psi(\hat x,\hat y)
\end{equation}
describes the complete action of the phase space operators $(\hat x_{i}^{q}, \hat p_{i}^{q})$ on $\mathcal{H}_{Q}$ satisfying the noncommutative Heisenberg algebra [2]. Now using the fact that $b|z\rangle=z|z\rangle$ we can introduce $|z,\bar z)=|z\rangle \langle z|$
as an upgraded version of the overcomplete coherent states in $\mathcal{H}_{Q}$ satisfying 
\begin{eqnarray}
b|z, \bar z) &=& z|z,\bar z).\nonumber\\
( z,\bar z |z^{\prime} z^{\prime})&=&tr_{C}(|z\rangle \langle z|z^{\prime}\rangle \langle z^{\prime}|)
=|\langle z|z\rangle |^{2}= e^{-|z-z^{\prime}|^{2}}
\end{eqnarray}
Following \cite{gracia}, we can now construct the `position' representation of a state $|\psi)=\psi(\hat x,\hat y)$ $\in \mathcal{H}_{Q}$ as
\begin{equation}
( z, \bar z |\psi)=tr_{C}(|z\rangle \langle z|\psi(\hat x,\hat y))=\langle z|\psi(\hat x,\hat y)|z\rangle
\end{equation}
which clearly corresponds to a lower symbol of the operator $\psi(\hat x,\hat y)$. In particular the position representation of the momentum eigenstates (\ref{delta}) turns out to be 
\begin{equation}
( z, \bar z|\vec p)=\sqrt{\frac{\theta}{2\pi}}e^{-\frac{\theta}{4}\vec p^{2}}e^{i\sqrt{\frac{\theta}{2}}(\bar z p+z\bar p)}; \,\, p=p_{1}+ip_{2}
\end{equation}
Using this one can easily show that 
\begin{equation}
\int{d^{2}p( z^{\prime}, \bar z^{\prime}|\vec p)( \vec p |z, \bar z)}=e^{-|z-z^{\prime}|^{2}}=( z^{\prime}, \bar z^{\prime}|z, \bar z)
\end{equation}
implying that $|\vec p)$ really forms a total family solving the identity in $\mathcal{H}_{Q}$ 
\begin{equation}
\int{d^{2}p \: |\vec p\:)( \vec p \:|}={\bf 1_{Q}}
\end{equation}
On the other hand 
\begin{equation}
\int{\frac{d\bar z dz}{\pi}(\vec p\:|z, \bar z)(z, \bar z|\vec{p}^{\:\prime}\:)}=e^{-\frac{\theta}{2}\vec p^{2}}\delta^{2}(\vec p-\vec p^{\:\prime})\ne \delta^{2}(\vec p-\vec p^{\:\prime})
\end{equation}
showing that the naive resolution of identity, the counterpart of (\ref{pscs12}) in $ \mathcal{H}_{Q}$, fails in this case 
\begin{equation}
\int{\frac{d\bar z dz}{\pi}( z, \bar z|z, \bar z)}\ne{\bf 1_{Q}}
\end{equation}
However, as we have mentioned in the preceding section, that lower symbols should be composed through the Voros star product (\cite{vo,bal}) as  
\begin{equation}
f(z,\bar z)\star g(z,\bar z) =f(z,\bar z) e^{\overleftarrow {\partial_{z}}\overrightarrow\partial_{\bar z}}g(z,\bar z)
\end{equation}
 Once done that, we can readily verify that 
\begin{equation}
\int{\frac{d \bar z dz}{\pi}(\vec p^{\:\prime}| z, \bar z)\star (z, \bar z|\vec p)}=\delta^{2}(\vec p-\vec p^{\:\prime})
\end{equation}
so that the appropriate resolution of identity in $\mathcal{H}_Q$ is given by 
\begin{equation}
\int{\frac{d\bar z dz}{\pi}| z, \bar z) \star ( z, \bar z|}={\bf 1_{Q}}
\end{equation}
Finally note that any element  $|\psi)=\psi(\vec x,\vec y)$ $\in$ $ \mathcal{H}_{Q}$ can be expanded in terms of the oblique operators $\Pi_{m,n}$ as,
\begin{equation}
 \psi(\vec x,\vec y)=\sum_{m.n}|m \rangle \langle m|\psi|n\rangle\langle n|\equiv \sum_{m,n}\psi_{m,n}\Pi_{m,n}
\end{equation}

implying that the oblique projectors $\{\Pi_{m,n}\}$ provide a complete set of states in $\mathcal{H}_{Q}$. Alternatively it follows from (\ref{oblproj}) that states $|z ,\bar z)$ $\in \mathcal{H}_{Q}$ also provide an overcomplete set if the coefficient involve the derivatives of Dirac's distribution function in ``position space'' which should also compose through Voros star product.

Extension of this analysis involving the formulation of NC Quantum Mechanics in 3D is rather non-trivial as the rotational invariance is broken in presence of such a constant(non-transforming) antisymmetric matrix $\theta_{ij}$ satisfying $[\hat x_{i},\hat x_{j}]=i\theta_{ij}$
as the dual vector $\vec \theta=\{\theta_{i}\equiv \frac{1}{2}\epsilon_{ijk}\theta_{jk}\}$ is pointed to a particular direction in space and which can only be restored by a twisted implementation of the rotation group (SO(3)) in a Hopf algebraic setting \cite{APBAK}. However one has to sacrifice the vectorial transformation property of the coordinate operators in D$\geq$3, which can now be identified as the primitive linear operators in a deformed Hopf algebra \cite{BCFT}. Further work in this direction is in progress and will be reported later.

\section{Concluding remarks}
In this paper, we have established that the CS  quantization map enables us to quantize singular classical functions, and, more generally,  distributions including  tempered distributions. More precisely, we are able to construct a reasonable and well behaved quantum angle and time operator for the free particle moving on the line. In particular our time operator is hermitian, verifies the canonical commutation relation with the Hamiltonian  up to order $\hbar$, and has the right semi-classical limit. Let us point out  the relevance of our work to the study  of a ``phase space formulation''  of quantum mechanics, which enables to mimic at the level of functions and distributions the  algebraic manipulations on operators  within the quantum context.  In particular, by carrying out  the CS quantization of Cartesian powers of planes, we could so have at our disposal an interesting ``functional portrait'' in terms of a ``star'' product on distributions for  the quantum logic based on manipulations of tensor products of quantum states. 

\section*{Acknowledgment}

One of us, BC would like to thank F. G. Scholtz and S. Vaidya for discussion.


\begin{thebibliography}{99}
\bibitem{timeop} Y. Aharonov and D. Bohm, Phys. Rev. 122 1649-1658 (1961); D.~H. Kobe and V.~C. Aguilera-Navarro  \emph{Derivation of the energy-time uncertainty relation}, Phys. Rev. A 50, 933-937 (1994); P. Pfeifer and J. Frohlich,  Rev. Mod. Phys. 67, 759 (1995); P. Busch \emph{The Time-Energy Uncertainty Relation}, arXiv:quant-ph/0105049 v2 10 Oct 2004.
\bibitem{szabo} R. Szabo, \emph{Quantum Field Theory on Noncommutative space}, Phys. Rep. 378, 207-299 (2003)
\bibitem{biswa} F.~G. Scholtz, B. Chakraborty, J. Govaerts and S. Vaidya, \emph{Spectrum of the noncommutative spherical well}, J. Phys. A 40
14581-14592 (2007)
\bibitem{pinzul} A. Pinzul and A. Stern, \emph{Edge states from defects on the noncommutative plane}, Mod. Phys. Lett. A 18, 2509-2516 (2003)
\bibitem{Sch26} E. Schr\"odinger, \emph{Der stetige \"Ubergang von der
Mikro- zur Makromechanik}, Naturwiss. 14, 664-666 (1926).
\bibitem{klau0} J.~R. Klauder, \emph{The Action Option and the Feynman Quantization of Spinor Fields in Terms of Ordinary 
c-Numbers}, Annals of Physics 11, 123-168 (1960); J.~R. Klauder, \emph{Continuous-Representation Theory I. Postulates of continuous-representation theory}, J. Math. Phys.  4, 1055-1058 (1963).
\bibitem{glauber0} R.~J. Glauber,  \emph{Photons correlations},  Phys. Rev. Lett. 10, 84-86 (1963).
\bibitem{sud} E.~C.~G. Sudarshan, \emph{Equivalence of Semiclassical and Quantum Mechanical Descriptions of Statistical Light Beams}, Phys. Rev. Letters  10, 277 (1963).
\bibitem{Berezin} F.~A. Berezin, \emph{General concept of quantization}, Commun. Math. Phys. 40, 153-174 (1975).
\bibitem{gracia} J.~M. Gracia-Bondia and J.~C. Varilly,  \emph{Algebra of distributions suitable for phase space quantum mechanics}, J. Math. Phys. 29, 869 (1998).
\bibitem{algen} S.~T. Ali  and M. Englis, \emph{Quantization methods: a guide for
physicists and analysts}, math-ph/0405065v1 (2004).
\bibitem{van} L. Van Hove, \emph{Sur le probl$\grave{\textrm{e}}$me des relations entre les transformations
unitaires de la M$\acute{\textrm{e}}$canique quantique et les transformations
canoniques de la M$\acute{\textrm{e}}$canique classique}, Bull. Acad. Roy. Belg., cl. des Sci. 37, 610-620 (1961).
\bibitem{kastrup} H.~A. Kastrup, \emph{A new look at the quantum mechanics of the harmonic oscillator},  Ann. Phys. (Leipzig) 7-8, 439-528 (2007).
\bibitem{wein} S. Weinberg, \emph{The cosmological constant problem}, Rev. Mod. Phys. 61, 1-23 (1989).
\bibitem{wig} E.~P. Wigner, \emph{Do the equations of motion determine the quantum mechanical commutation relations?}, Phys. Rev. 77, 711 (1950).
\bibitem{watson} G.~N. Watson, \emph{A treatise on the theory of Bessel functions}, Cambridge Mathematical Library (1995).
\bibitem{dirac27} P.~A.~M. Dirac  Proc. R. Soc.  London Ser. A  114, 243-265 (1927).
\bibitem{pauli58} W. Pauli,  Die allgemeinen Prinzipien der Wellenmechanik, Hanbuck
der Physik, vol. 1, Springer Verlag, Berlin (1958).

\bibitem{galapon02} E. Galapon, Proc. R. Soc. Lond. A  458, 451-472 (2002).
\bibitem{giann02} R. Giannitrapani, [arXiv: quant-ph/0302056] (2002).
\bibitem{toller96}  M. Toller, gr-qc/9605052 v1 (1996).
\bibitem{bog} P. Boggiatto, E. Cordero \emph{Anti-Wick quantization of tempered distributions},  Progress in analysis, Vol. I, II, Berlin (2001), 655-662, World Sci. Publ., River Edge, NJ (2003).
\bibitem{bogg} P. Boggiatto, E. Cordero, K. Gr$\ddot{\textrm{o}}$chenig \emph{Generalized Anti-Wick Operators with Symbols
in Distributional Sobolev spaces}, Integral Equations Operator Theory 48, no. 4, 427-442 (2004).
\bibitem{schwartz} L. Schwartz, \emph{M$\acute{\textrm{e}}$thodes math$\acute{\textrm{e}}$matiques pour les sciences physiques}, Hermann (1961).
\bibitem{hir} A.~C. Hirshfeld, P. Henselder, \emph{Deformation Quantization in the Teaching of Quantum Mechanics}, quant-ph/0208163  (2002).
\bibitem{vo} A. Voros, \emph{Wentzel-Kramers-Brillouin method in the Bargman representation}, Phys. Rev. A 40, 6814-6825 (2002).
\bibitem{da} M. Daoud, \emph{Extended Voros product in the coherent states framework}, Phys. Lett. A 309, 167-175 (2003).
\bibitem{alex} G. Alexanian, A. Pinzul, A. Stern, \emph{Generalized coherent state approach to star products and applications to the fuzzy sphere}, Nuc. Phys. B 600, 531-547 (2001).
\bibitem{bal} A.P.Balachandran, S.Kurkcuoglu and S.Vaidya, \emph{Lectures on Fuzzy and Fuzzy SUSY Physics}, World Scientific, Singapore (2007).
\bibitem{kl} J.~R. Klauder, \emph{Fundamentals of Quantum Optics},  (1968).
\bibitem{ca} K.~E. Cahill, \emph{Coherent-State Representations for the Photon Density}, Phys. Rev. 138, B1566 (1965).
\bibitem{mi} M.~M. Miller, \emph{Convergence of the Sudarshan Expansion for the Diagonal Coherent-State Weight Functional}, J. Math. Phys. 9, 1270 (1968).
\bibitem{glauber1} R.~J. Glauber, \emph{The quantum theory of optical coherence},  Phys. Rev. 130, 2529 (1963).
\bibitem{main:ch1:glauber2} R.~J. Glauber, \emph{Coherent and incoherent states of radiation field}, Phys. Rev. 131, 2766 (1963).
\bibitem{APBAK} See for example the recent review: E. Akofor, A.~P. Balachandran, A. Joseph,\emph{Quantum Fields on the Groenewold-Moyal Plane}, Int. J. Mod. Phys. A23:1637-1677 (2008) and the references therein.
\bibitem{BCFT} P.G Castro, B. Chakraborty, and F. Toppan, \emph{Wigner oscillator, twisted Hopf algebras, and second quantization}, J. Math. Phys. 49 082106 (2008)
\end{thebibliography}
\end{document}